\documentstyle[12pt,preprint,aps]{revtex}

\begin{document}
%\draft

\title{Biot-Savart-like law in electrostatics}

\author{M\'ario H. Oliveira and Jos\'e A.Miranda\footnote{Corresponding author\\e-mail:jme@lftc.ufpe.br}}
\address{Laborat\'{o}rio de F\'{\i}sica Te\'{o}rica e Computacional,
Departamento de F\'{\i}sica,\\ Universidade Federal de Pernambuco, 
Recife, PE  50670-901 Brazil}
\date{\today}
\maketitle

\begin{abstract} 
The Biot-Savart law is a well-known and powerful theoretical tool used to calculate magnetic fields due to currents in magnetostatics. We extend 
the range of applicability and the formal structure of the Biot-Savart 
law to electrostatics by deriving a Biot-Savart-like law suitable for calculating electric fields. We show that, under certain circumstances, 
the traditional Dirichlet problem can be mapped onto a much simpler Biot-Savart-like problem. We find an integral expression for 
the electric field due to an arbitrarily shaped, planar region 
kept at a fixed electric potential, in an otherwise grounded plane. 
As a by-product we present a very simple formula to compute the 
field produced in the plane defined by such a region. We illustrate 
the usefulness of our approach by calculating the electric field produced by 
planar regions of a few nontrivial shapes.
\end{abstract}

\section{Introduction}
The Biot-Savart law~\cite{Halliday,Griffiths,Jackson,Eyges} is one of the most 
basic relations in electricity and magnetism. It allows one to 
determine the total magnetic field ${\bf B}$ at a given point in 
space as the superposition of infinitesimal contributions ${\bf {\rm \bf d}B}$, 
caused by the flow of current $I$ through an infinitesimal 
path segment ${\bf {\rm \bf d}s}$, oriented in the same direction as 
the current. If ${\bf \hat{r}}$ is the position unit-vector pointing from the 
element of length to the observation point $P$, then the total field 
contribution due to a closed circuit $C$ at $P$ is given by  a closed 
{\it line} integral~\cite{Halliday,Griffiths,Jackson,Eyges}
\begin{equation}
\label{newBiot}
{\bf B}=\frac{\mu_{0}I}{4\pi} \oint_{C} \frac{({\bf {\rm \bf d}s} \times {\bf \hat{r}})}{r^{2}}, 
\end{equation}
where $\mu_{0}$ is the permeability of free space and the path of 
integration is along the wire. 

The usefulness of the Biot-Savart law 
goes far beyond its basic textbook applications. Equation~(\ref{newBiot}) 
is a much studied integral form which arises in various interesting 
physical problems involving topological defects~\cite{Kroener}, 
magnetic fluids~\cite{Jackson2}, amphiphilic monolayers~\cite{Kessler}, 
type-I superconductors~\cite{Goldstein} and the $n$-body 
problem of celestial mechanics~\cite{Buck}. In the 
framework of classical electrodynamics the Biot-Savart law 
provides a useful technique to calculate the magnetic 
field generated by current carrying wires. 
Equation~(\ref{newBiot}) is quite general, valid for nonplanar 
current loops of arbitrary shape. Obviously, the presence 
of a cross-product in the formula may introduce some difficulties 
in practical calculations. Ultimately, the range of problems to 
which equation~(\ref{newBiot}) can be applied is limited primarily 
by the difficulty experienced in performing the integrations. These 
difficulties are less serious if the loop $C$ is flat 
and if the observation point lies in the plane of the loop. It has 
been recently shown~\cite{Mir3} that the magnitude of the magnetic field 
due to an arbitrarily shaped (not self-intersecting), planar, current 
carrying wire at a point lying on the wire's plane can be written as
\begin{equation}
\label{result}
 B=\frac{\mu_{0}I}{4\pi} \oint_{C} \frac{d\theta}{r}. 
\end{equation}
Equation~(\ref{result}) is very simple and compact, expressing $B$ 
in terms of the wire shape $r=r(\theta)$, where $\theta$ denotes 
the polar angle. As shown in reference~\cite{Mir3}, this basic result 
expands the degree of applicability of the Biot-Savart law 
allowing exact, closed form solutions for a whole new set 
of elementary problems. 

Based on the general nature of the Biot-Savart law, its connection 
to recent physics research topics~\cite{Jackson2,Kessler,Goldstein,Buck}, 
and its success in performing magnetic field calcultions, we 
felt motivated to investigate the following question: is it possible 
to formulate a Biot-Savart-like law in electrostatics ? If so, 
what sort of electrostatic problem could be more easily solved 
by such an approach? In this work we address these issues and 
show, for the first time, that it is indeed possible to 
formulate a Biot-Savart-type law in the realm of electrostatics. 

The formulation we propose is suitable to calculate the electric field 
due to an arbitrary shaped, planar region maintained at 
a fixed scalar potencial V, with the rest of the plane held at 
zero potential. We show that the calculation of the electric field 
produced by such a region is analogous to the evaluation of the 
magnetic field due to a flat current carrying wire of the same 
shape (figure 1). We point out that the ``brute-force" 
calculation of the electric field due to such an arbitrarily shaped 
region, kept at a fixed potential, by directly applying conventional boundary-value techniques~\cite{Extra1} may be quite challenging. 
The nontrivial nature of the problem comes from the fact that 
we do not know the surface charge distribution in advance. 
Our eletrostatic Biot-Savart law 
provides a much simpler way to perform this nontrivial 
electric field calculation, and allows one to borrow many of the 
standard procedures ordinarily used in the corresponding 
magnetostatic situation. Therefore, complicated electrostatic 
problems may have straightforward solutions, if 
solved by the Biot-Savart-like approach we develop here. 
We stress that, even though our derivation and new 
results~(\ref{EBiot}) and~(\ref{EBiot2}) presented below are quite 
simple, they have not been derived in any standard electromagnetism 
book or journal publication. This work comes to fill this 
gap, offering a new and simple tool to perform electric 
field calculations.

As a by-product, we obtain a very compact expression 
for the electric field in the plane of the flat sheet 
(see equation~(\ref{EBiot2})), which is as 
simple as its magnetic field counterpart given by 
equation~(\ref{result}). Finally, we illustrate our results by 
explicitly calculating the electric field produced by a constant 
potential, flat region that has the shape of a regular, $n$-sided 
polygon, for observation points located along its axis of 
symmetry. The calculation for the field produced in the plane 
of regions having other peculiar shapes (elliptical, spiral, regularly 
undulating) is also presented.

\section{Electric field calculation}

Consider an arbitrarily shaped, two-dimensional region, 
located in the $x$-$y$ plane, kept at a fixed potential V while 
the rest of the plane is held at zero potential (see figure 2). 
The shape of the flat plate is determined by a closed boundary curve $C$. 
We want to calculate the electric field caused by this charge 
configuration at a given observation point P located by the vector 
${\bf x}$ 
\begin{equation}
\label{Efield}
{\bf E}({\bf x})=- {\bf \nabla} \Phi({\bf x}), 
\end{equation}
where $\Phi({\bf x})$ denotes the electric scalar potential. 

Since the scalar potential is specified everywhere on the $x$-$y$ plane, 
we have to solve a Dirichlet problem~\cite{Extra1}. We apply Dirichlet boundary 
conditions, i.e., $G_{D}({\bf x}, {\bf x'})~|_{z'=0}=0$, where 
$G_{D}({\bf x}, {\bf x'})$ denotes the Green's function for Dirichlet conditions, and the primed vector ${\bf x'}$ locates the charge 
distribution points. Using Green's theorem and the fact that the volume 
charge density $\rho({\bf x'})=0$, the electrostatic potential 
for $z>0$ can be expressed in terms of the value of the 
potential on the plate $\Phi({\bf x'})=V$~\cite{Extra1}
\begin{equation}
\label{Phi}
\Phi({\bf x})=-\frac{1}{4 \pi} \int \Phi({\bf x'}) \left ( \frac{\partial{G_{D}}}{\partial {n'}} \right )~{\rm d a'},
\end{equation}
where ${\bf \hat {n'}}={-\bf  \hat{z'}}$ is the outward unit normal 
and ${\rm d a'}$ is an infinitesimal area element of the plate.

By employing the method of images~\cite{Extra1}, 
we can easily find the Green's function by considering the potential 
in ${\bf x}$ due to a unit point charge located at a point in the 
region $z>0$, plus the potential of an image charge placed 
in a symmetric position in the lower-half plane $z<0$. 
Through this procedure we calculate 
$(\partial{G_{D}}/ \partial {n'})=(-\partial{G_{D}} / \partial {z'}) |_{z'=0}$ 
explicitly, and rewrite equation~(\ref{Phi}) as 
\begin{equation}
\label{Phi2}
\Phi({\bf x})=\frac{V}{2 \pi}~\Omega({\bf x}),
\end{equation}
where
\begin{equation}
\label{Solid}
\Omega({\bf x})=\int \frac{ {(-\bf \hat{z} )} \cdot ({\bf x'} - {\bf x})}{|{\bf x} - {\bf x'}|^{3}}~{\rm d a'}
\end{equation}
is the solid angle subtended by the surface (flat plate of area $a'$) 
spanning the loop $C$ as seen from ${\bf x}$. The sign convention 
established for the solid angle is the following: $\Omega({\bf x})$ 
is positive when the observation point $P$ views the ``inner" side of 
the flat plate, in other words, $\Omega({\bf x})>0$ if the unit normal 
points away from $P$. This sign convention is the same as the one 
adopted in references~\cite{Jackson,Eyges}.

By substituting equations~(\ref{Phi2}) and~(\ref{Solid}) into the 
electric field equation~(\ref{Efield}) and using the well-known relation~\cite{Extra2}
\begin{equation}
\label{Solid2}
{\bf \nabla} \Omega({\bf x})=\oint_{C} \frac{ {\bf {\rm \bf d}s'}  \times ({\bf x} - {\bf x'})}{|{\bf x} - {\bf x'}|^{3}},
\end{equation}
we finally obtain
\begin{equation}
\label{EBiot}
{\bf E}({\bf x})=\frac{V}{2 \pi} \oint_{C}  \frac{({\bf x} - {\bf x'}) \times  {\bf {\rm \bf d}s'}}{|{\bf x} - {\bf x'}|^{3}},
\end{equation}
where ${\bf {\rm \bf d}s'}$ is an element of length of the 
integration path $C$. The direction of the 
integration around $C$ is determined by the direction of the outward 
unit normal via the right-hand rule. Notice that the cross-product in~(\ref{EBiot}) has a reversed order in comparison to the one 
that appears in the magnetostatic case~(\ref{newBiot}). This 
distinction is necessary to give the correct direction for the 
electric field vector at the observation point $P$. 

Equation~(\ref{EBiot}) is our central result. It allows the 
calculation of the electric field through a Biot-Savart-type law. 
Notice that~(\ref{EBiot}) is written in terms of a line 
integral, so to calculate the electric field we just need to take 
into account the contributions coming from the boundary 
contour $C$. This result makes the solution of the electrostatic 
problem much easier. Under the circumstances studied 
in this work, the traditional Dirichlet problem can be 
mapped onto a Biot-Savart-like problem, a mapping that can 
simplify considerably the computation of the electric field 
in many problems of interest.

As it was in the magnetic case~\cite{Mir3}, a much simpler expression for the 
field can be obtained if we concentrate our attention in the 
calculation of the electric field at observation points that lie 
in the plane of the charge distribution. Using the same arguments 
as those presented in reference~\cite{Mir3} it can be easily shown that
\begin{equation}
\label{EBiot2}
{\bf E}({\bf x})=\frac{V}{2 \pi} \oint_{C} \frac{d\theta}{r} ~{\bf \hat{z}},
\end{equation}
where $r=|{\bf x} - {\bf x'}|$. Equation~(\ref{EBiot2}) is a surprisingly 
simple result. Note that this line integral 
expression~(\ref{EBiot2}) works for flat charge distributions 
of any boundary shape, including those described by curves $C$ 
which are not single valued functions of $\theta$. In addition, 
it is valid for observation points $P$ located either inside 
or outside the loop $C$. To illustrate our approach, in the 
next section we calculate the electric field due to specific 
planar charge configurations of some representative shapes.

\section{Illustrative examples}

In this section we illustrate the usefuness the the Biot-Savart-like law~(\ref{EBiot}) by discussing a class of electrostatic problem that is 
trivially solved by using~(\ref{EBiot}) or~(\ref{EBiot2}), 
but that would require a much 
more involved solution otherwise. We start by applying 
equation~(\ref{EBiot}) to calculate the electric field due 
to a thin, flat plate, which has the shape of a regular 
n-sided polygon, inscribed in a circle of radius $a$ (see figure 3). 
Consider that the flat plate is located in the $x$-$y$ plane with its 
center at the origin,  and that it is maintained at a 
fixed potential $V$. In the plane $z=0$ the region outside 
the plate is held at zero potential. We wish to 
compute the electric field at a point $P$ in the $z$-axis.

As discussed in section 2, by inspecting equation~(\ref{EBiot}) 
we notice that in order to calculate the electric 
field due to this flat configuration, we just need to take into
account the contributions coming from the boundary contour $C$. Let 
us analyze this point a little more carefully: the total 
electric field at point $P$ in the $z$-axis can be written as 
the superposition of the field due to $n$ triangles obtained by 
joining the center of circle to the vertices of the polygon (figure 3). 
If we sum all the contributions from these various triangular paths, 
we will be left with the integration around the contour $C$. 
This happens because the sense of integration along their 
common sides is opposite for two adjacent triangles, making the 
contributions from the common sides to cancel. Therefore, 
after performimg the integration around all the triangles, the only 
nonzero contributions come from the $n$ straight edges that 
define the plate's boundary contour $C$. With these considerations 
in mind we turn to the field calculation itself.

When the field contributions of the $n$ sides are summed vectorially, 
the horizontal ($x$-$y$ plane) components add to zero. By symmetry, 
only the vertical components located along the $z$-axis will survive. 
From figure 3 we verify that the total electric field at $P$ 
is given by
\begin{equation}
\label{Etotal}
E=n~E_{n}\cos{\alpha},
\end{equation}
where $E_{n}$ is the net field due to just one of the $n$ triangles 
that compose the polygon and 
$\cos{\alpha}=a \cos{(\pi/n)}/\sqrt{a^2\cos^{2}{(\pi/n)} + z^{2}}$. 
Considering the fact that the only nonzero contribution from such 
triangular path comes from the single edge of the polygon, we 
employ equation~(\ref{EBiot}) to get
\begin{equation}
\label{nsided2}
E_{n}=\frac{V}{\pi}~ \frac{a \sin{(\pi/n)}}{ \sqrt{ \left [ a^2\cos^{2}{(\pi/n)} + z^{2} \right ]~\left [ a^{2} + z^{2} \right ]}}.
\end{equation}

Using equations~(\ref{Etotal}) and~(\ref{nsided2}) we obtain the 
total electric field along the $z$-axis
\begin{equation}
\label{nsided}
{\bf E}=\frac{V}{2 \pi}~\left \{ \frac{na^{2}\sin{(2\pi/n)}}{\left [ a^2\cos^{2}{(\pi/n)} + z^{2} \right ]~\sqrt{a^{2} + z^{2}}} \right \} ~{\bf \hat{z}}.
\end{equation}
This result is quite handy since it provides, all at once, 
the calculation of the electric field along the $z$-axis 
due to any regular polygon of $n$ sides (equilateral triangle, 
square, pentagonal, etc.). We point out that our {\it electric field} result~(\ref{nsided}) agrees with the equivalent (but 
conceptually distinct) formula for the {\it magnetic field} 
at the axis of an n-sized polygonal, current carrying circuit, 
previously obtained in reference~\cite{Chirgwin}.

It is worth mentioning that result~(\ref{nsided}) can also be used 
compute the magnitude of the electric field along the axis passing 
through the center of a {\it circular} conducting plate of radius $a$. 
This can be easily obtained by taking the 
limit $n \rightarrow \infty$ in equation~(\ref{nsided}) yielding
\begin{equation}
\label{circle2}
{\bf E}=\frac{Va^{2}}{[a^{2} + z^{2}]^{3/2}}~{\bf \hat{z}}. 
\end{equation}
This limit agrees with reference~\cite{Jackson3}, in which 
the potential calculation is done solely for the circular plate, 
by employing traditional boundary-value problem techniques.

We conclude this section by calling the attention of the reader 
to the fact that the calculation of the electric field at 
observation points lying in the plane of flat plates of various 
shapes can be obtained, with great facility, by directly using 
our expression~(\ref{EBiot2}). For example, the magnitude 
of the electric field in the plane of elliptical, spiral 
shaped and harmonically deformed circular plates (see figure 4), 
kept at a fixed potential $V$, are readily obtained by ~(\ref{EBiot2}).
Table 1 displays the values of the electric field at observation 
points $P$, for these three characteristic flat regions.  So, 
despite the boundary curve especific geometry, the $E$ field 
calculation may be promptly performed by using our Biot-Savart-like 
approach. In practical terms the difficulty of having a complicated 
boundary shape is not a very serious obstacle in order to compute $E$ 
in closed form, as long as the related integrations are not terribly 
hard to handle.

\section{Concluding Remarks}

In this work we show that, under certain circumstances, complicated 
electrostatic problems may have straightforward solutions, if 
solved by a Biot-Savart-like approach. We consider the general situation 
in which a flat region of arbitrary shape is kept at a fixed 
potential $V$, in a otherwise grounded plane. The objective is to compute the 
electric field due to this nontrivial charge distribution at a given point in 
space. At first glance, such a calculation looks very involved, mainly 
because we have no prior knowledge about the precise charge distribution 
on the planar region. We derived a Biot-Savart-like law suitable to deal 
with such electrostatic situation, allowing the calculation of the 
electric field to be done in a simple fashion. 

This Biot-Savart-like law is written in terms of a line integral 
along the boundary contour defined by the nonzero potential region, 
so that the calculation does not require information about the charge distribution in the bulk. As a by-product, we show that for observation points 
located in the $z=0$ plane, the Biot-Savart-like expression can be written 
in a very simple and compact form, as was the case for similar magnetic 
field calculations recently published in reference~\cite{Mir3}. 
We illustrate our results by calculating the electric field along the 
axis of a polygonal flat region of $n$ sides, kept at a constant 
potential $V$. In addition, we calculate the magnitude of the electric 
field in the plane of charged plates presenting nontrivial boundary 
geometries such as elliptical, spiral and regularly undulating borders. 
In summary, we show that it is possible to define a Biot-Savart-like 
in electrostatics. Our new approach provides an alternative and 
simple technique to solve a class of complicated boundary value 
problems in electrostatics.

\vspace{0.5 cm}
\begin{center}
{\bf ACKNOWLEDGMENTS}
\end {center}
\noindent
This work was supported by CNPq and FINEP (Brazilian Agencies).

\pagebreak

\noindent
{\Large {\bf Figure Captions}}

\vskip 0.5 in
\noindent
{\bf Figure 1:} Two equivalent systems: (a) arbitrarily shaped, planar wire 
carryng a steady current $I$, and (b) a flat charge distribution of 
the same shape, maintained at a fixed potential $V$, with zero potential 
in the rest of the plane. In (a) the magnetic field ${\bf B}$ 
can be calculated with the help of the usual Biot-Savart 
law (equation ~(\ref{newBiot})), while in (b) the calculation of 
the electric field ${\bf E}$ may be performed with the help of 
an electrostatic Biot-Savart-like law (equation~(\ref{EBiot})).
\vskip 0.25 in
\noindent
{\bf Figure 2:} Charged flat plate located in the $x$-$y$ plane, bounded 
by the curve $C$ and kept at a fixed potential $V$. The potential is 
set to zero in the region of the plane $z=0$ outside the curve $C$.
Vector ${\bf x'}$ locates the source point, while vector ${\bf x}$ 
refers to the field point. The observation point is denoted by $P$.
\vskip 0.25 in
\noindent
{\bf Figure 3:} Schematic view of a regular $n$-sized polygonal plate, 
kept at a potential $V$. The plate lies in the $x$-$y$ plane, while the observation point $P$ is located along the $z$-axis. Each individual 
triangle defines the angle $2\pi/n$. The direction 
of integration around $C$ is determined by the outward unit-normal 
${\bf \hat {n}}={-\bf  \hat{z}}$ via the right-hand rule.
\vskip 0.25 in
\noindent
{\bf Figure 4:}(a) Elliptical plate, presenting major 
axis $2a$ and minor axis $2b$, centered at point $P$; (b) Harmonically 
deformed plate (Perturbed Circle) $r(\theta)$=$a$ $[1 + \epsilon \cos(n \theta)]$ , centered at $P$, for $n=5$ and $\epsilon=0.5$; (c) Logarithmic 
spiral plate. The magnitude of the electric fields at observation 
point $P$ due to these shapes are listed in Table 1.

\pagebreak

\begin{table}[ht] 
\begin{centering}
\begin{tabular}{|c|c|c|}
\hline
{\em\bf Boundary Shape} &{\em\bf Parametric Equation } &{\em\bf Electric Field Magnitude} \\ \hline \hline
{$Elliptical$} & $r(\theta)=\frac{ab}{\sqrt{a^{2}\sin^{2}\theta + b^{2}\cos^{2}\theta}}$ & $E=\frac{2V}{\pi a}~{\em\cal G} \left ( \sqrt{1 - \frac{a^{2}}{b^{2}}} \right )$ \\ 
& &\\  \hline \hline 
{$Perturbed ~Circle$} &$r(\theta)= a[1 + \epsilon \cos{(n \theta)}]$ &$E=\frac{V}{a \sqrt{1 - \epsilon^{2}}}$ \\ 
& & \\ \hline \hline 
{$Spiral$} &$r(\theta)=q e^{p \theta}$ &$E=\frac{V}{2 \pi q} \left[ \frac{1 - e^{-2 \pi p}}{p} \right]$ \\ 
& & \\ \hline
\end{tabular}
{\caption{\protect Electric field in the plane of flat regions presenting nontrivial boundary shapes. Note: the function ${\cal G}$ denotes the complete elliptic integral of the second kind \protect\cite{Grad}.}}
\end{centering}
\end{table}

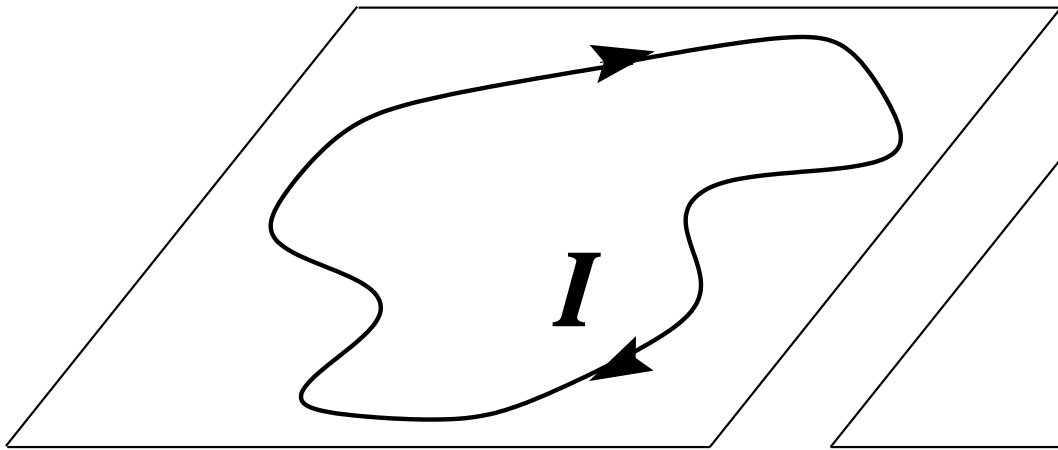
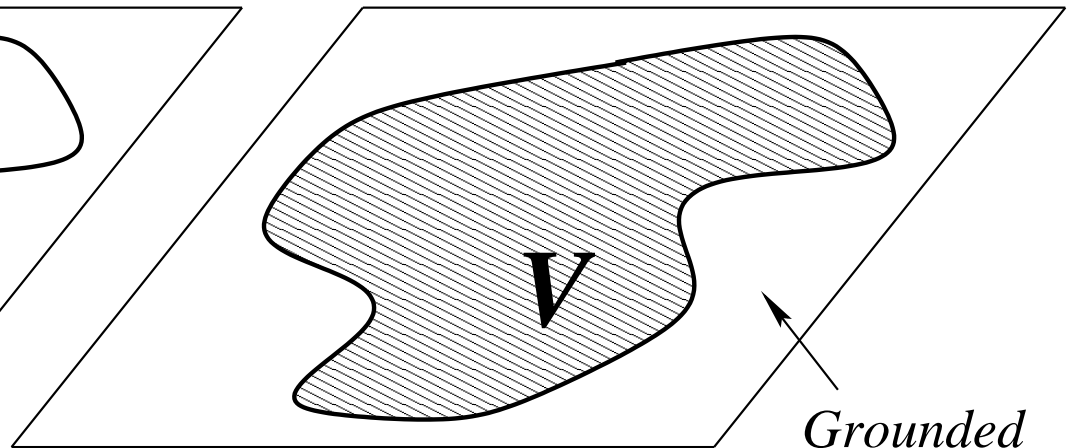
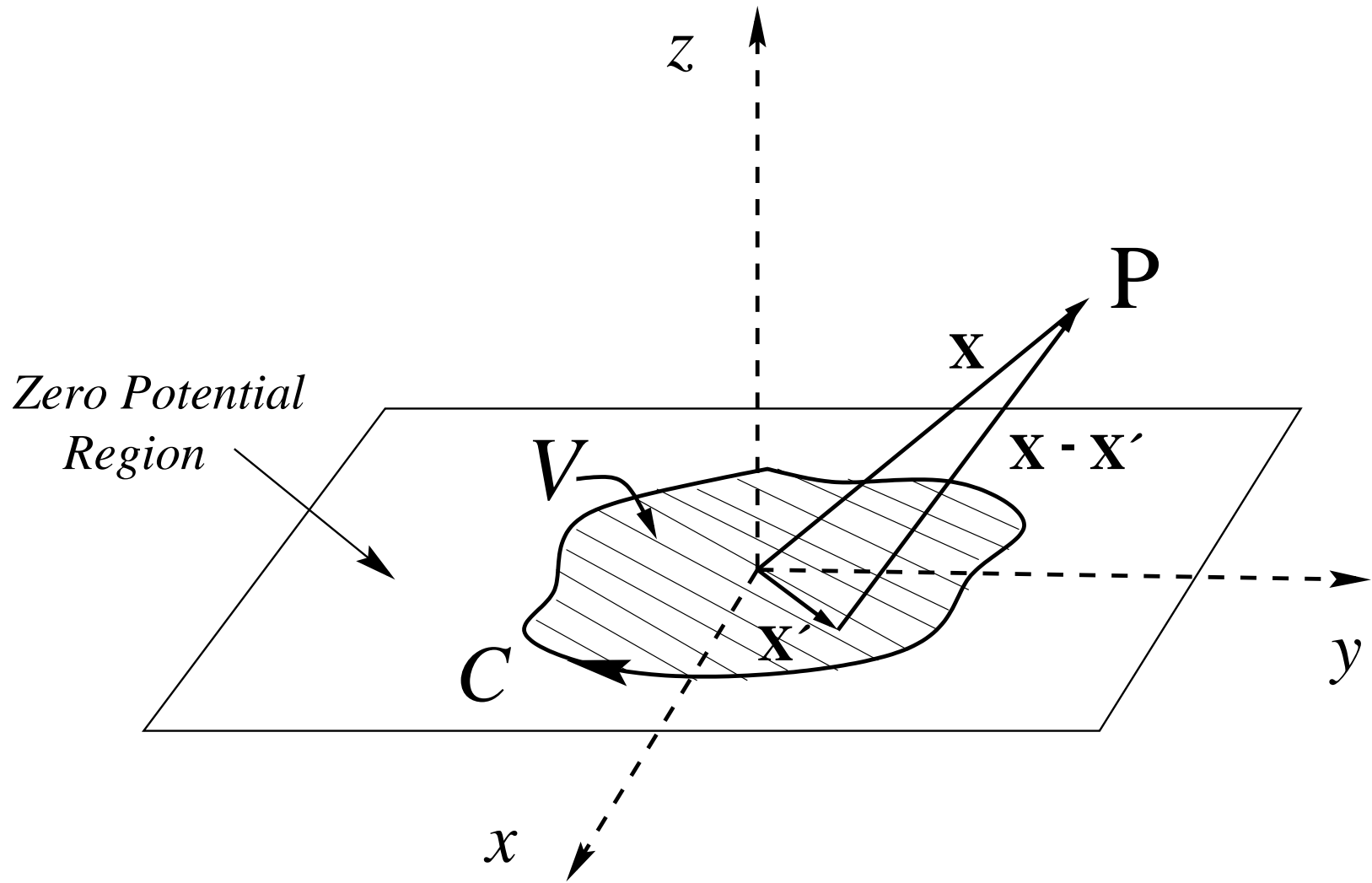
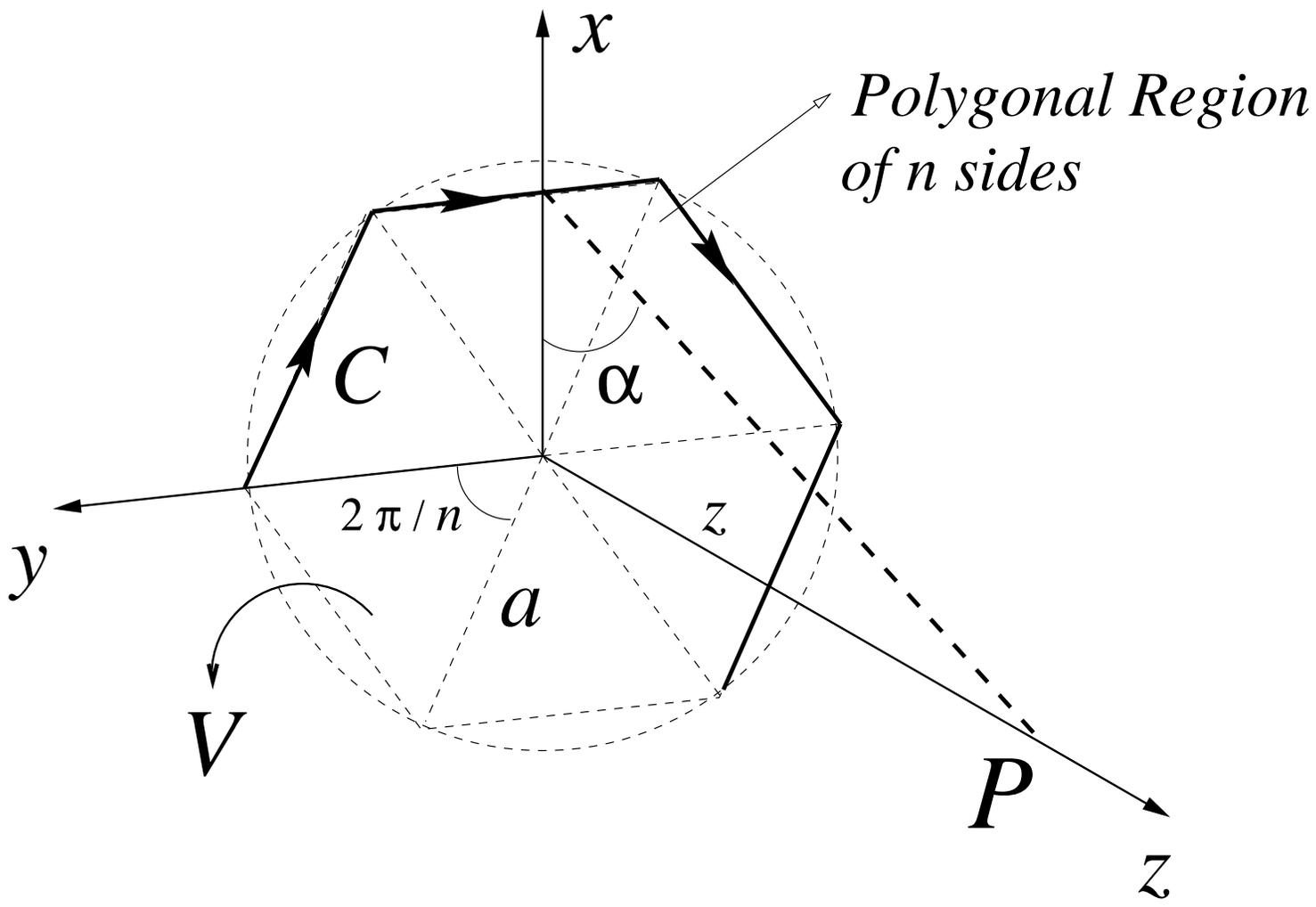
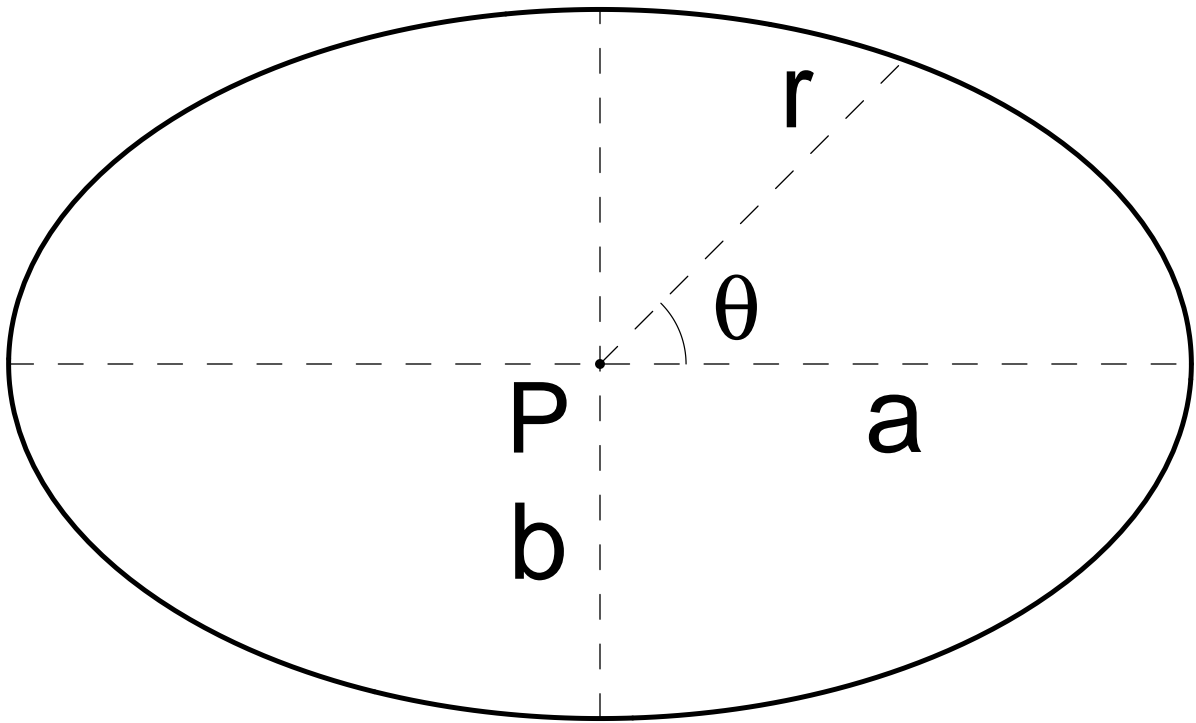
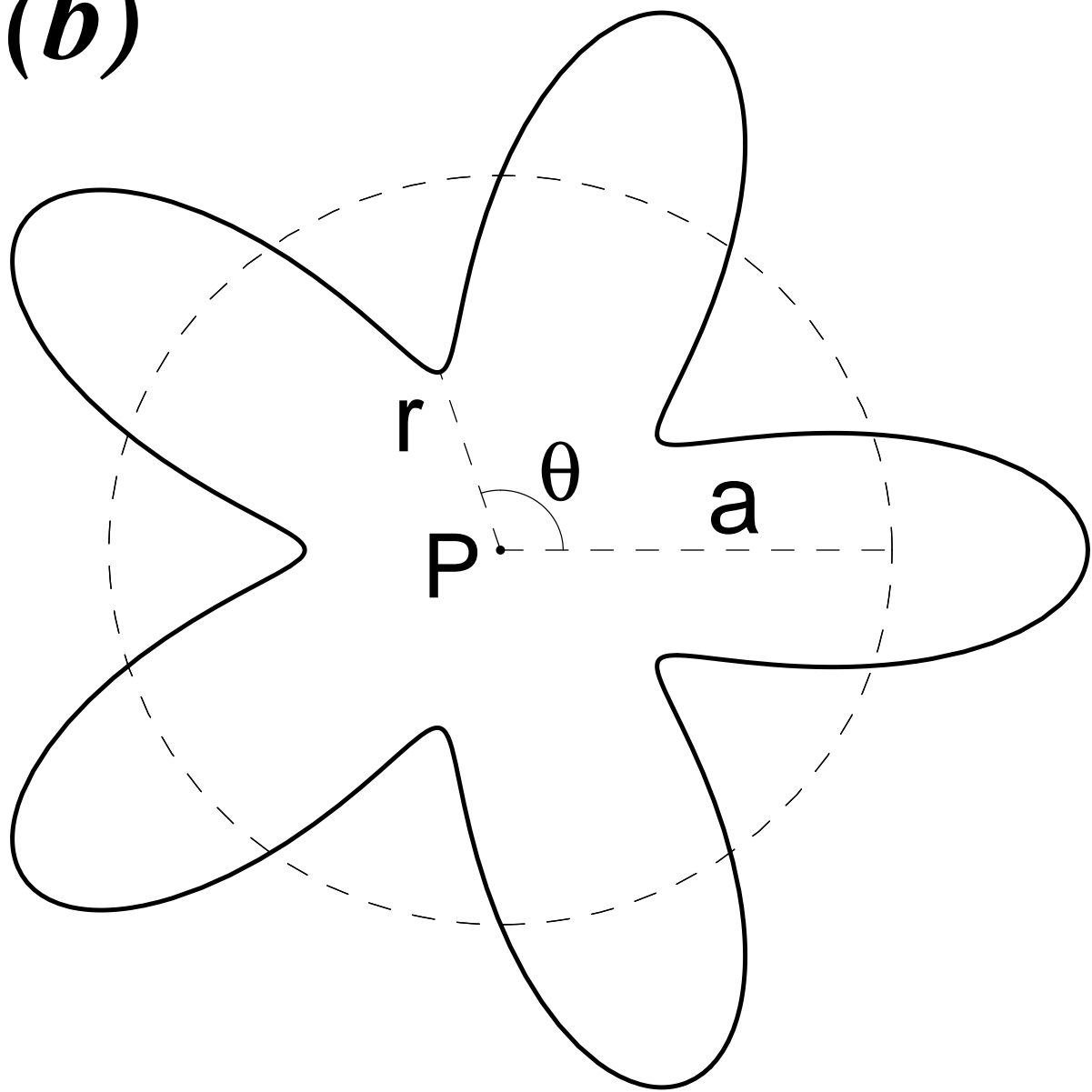
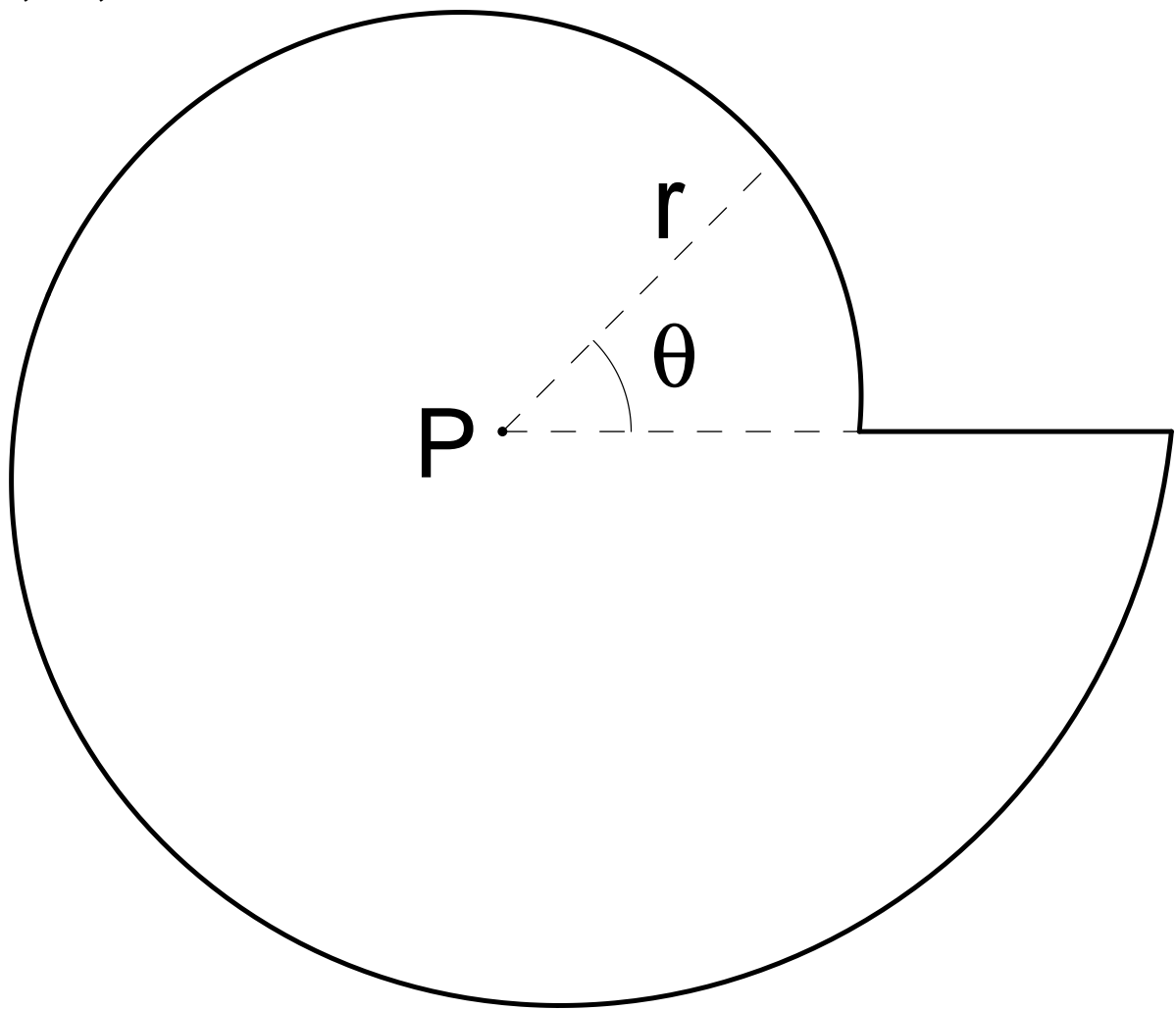
\end{document}